\documentstyle[twocolumn,osa,aps]{revtex}
\input epsf
\newcommand\inv{^{-1}}
\newcommand{\ZZ}{Z\kern-.5em{Z}}

\newcommand{\cE}{{\cal E}}
\newcommand{\bK}{{\bf K}}
\newcommand{\bX}{{\bf X}}
\newcommand{\bR}{{\bf R}}
\newcommand{\bT}{{\bf T}}
\newcommand{\bA}{{\bf A}}
\newcommand{\bB}{{\bf B}}
\newcommand{\bU}{{\bf U}}
\newcommand{\bV}{{\bf V}}
\newcommand{\cN}{{\cal N}}
\newcommand{\cU}{{\cal U}}
\newcommand{\cV}{{\cal V}}
\newcommand{\bc}{{\bf c}}
\newcommand{\br}{{\bf r}}
\newcommand{\bt}{{\bf t}}
\newcommand{\tbr}{{\tilde{\bf r}}}
\newcommand{\tbt}{{\tilde{\bf t}}}
\newcommand{\la}{{\langle}}
\newcommand{\ra}{{\rangle}}
\newcommand{\cJ}{{\aro{\cal J}}}

\newcommand{\beq}[1]{ \begin{equation}\label{#1}}
\newcommand{\eeq}{\end{equation}}
\newcommand{\ba}{\begin{array}}
\newcommand{\ea}{\end{array}}
\newcommand{\beqa}{\begin{eqnarray}}
\newcommand{\eeqa}{\end{eqnarray}}
\newcommand{\ds}{\displaystyle}

\newcommand{\aro}{\overrightarrow}

\newcommand{\hf}{{\textstyle \frac{1}{2}}}

\newcommand{\R}{I\kern-.3em{R}}
\begin{document}
\title{A NUMERICAL STUDY OF ABSORPTION BY MULTILAYERED BIPERIODIC STRUCTURES}
\author{G. Berginc\thanks{Thomson CSF-Optronique, rue Guynemer, BP 55, 78283 Guyancourt cedex, France.},
C. Bourrely\thanks{Birkbeck College, Physics Department, University of London,
London WC1E, UK. Permanent address: CPT, CNRS-Luminy, Marseille.},
C. Ordenovic$^{*}$ and B. Torr\'esani}
\address{CPT, CNRS-Luminy, case 907, 13288 Marseille Cedex 09, France}
\date{May 1997}
\maketitle
\begin{abstract}
We study the electromagnetic scattering by multilayered biperiodic
aggregates of dielectric layers and gratings of conducting plates.
We show that the characteristic lengths of such structures provide
a good control of absorption bands. The influence of the physical
parameters of the problem (sizes, impedances) is discussed.
\end{abstract}
\section{INTRODUCTION}
\label{se:intro}
Electromagnetic absorbers and frequency selective surfaces (FFS for short)
have recently received an increasing interest. There is a growing need for
electromagnetic absorbers, and in particular for lighter, thinner
and more highly absorbing materials. Frequency selective surfaces are
generally made of planar screens with periodic or biperiodic metallizations.
One generally considers two types of FFS: {\em capacitive FFS} are
transparent at low frequencies; {\em inductive FFS} are reflecting ones.
Their behavior at the resonance frequency is complementary.
Capacitive FFS consist of arrays of metal patches embedded in
a dielectric structure, which may be a stratified one. The dielectric
structure provides the mechanical support of the FFS. Inductive FFS
consist of perforated screens.

Such frequency selective surfaces have been considered by several authors~\cite{chen,mittra,rub.ber} who have proposed various
approaches for the numerical resolution of the corresponding
scattering problem. Efficient numerical methods are
now available for the analysis and design of FFS, as we shall show.

The purpose of this paper is to show that the absorption bands of such structures may be controlled by combining the performances of capacitive
FFS and electromagnetic absorbers. To vary the frequency response of a
FFS, the standard method consists in varying the geometry of the array 
elements. We give efficient computational methods for analyzing this
kind of structure. The representation of the transmitted and reflected
fields is obtained by applying resistive boundary conditions
to include a general surface impedance in the problem formulation.

We provide examples of such periodic or
biperiodic structures, whose absorption bands can easily
be controlled by varying some of the characteristic
lengths of the system. More precisely, we consider
multilayers made of dielectric
stacks and surface gratings with various shapes and
sizes. We show that such structures yield absorption bands,
and that the location and bandwidth of such bands may be controlled
by varying the characteristic sizes of the structure.

This paper is organized as follows. After this introduction, we describe
in Section~\ref{se:model} the details of the diffracting structures we
consider, and the model we use to solve numerically the corresponding
diffraction problem. Then we develop in section~\ref{se:num} the
numerical resolution method, and discuss a series of examples.
Finally, section~\ref{se:concl} is devoted to the conclusions.
More technical aspects concerning the mathematical background and
numerical details are discussed in three appendices at the
end of this paper.

\section{MODELLING THE BIPE\-RIODIC STRUCTURES}
\label{se:model}
\begin{figure}
\epsfxsize=8cm
\centerline{\epsfbox{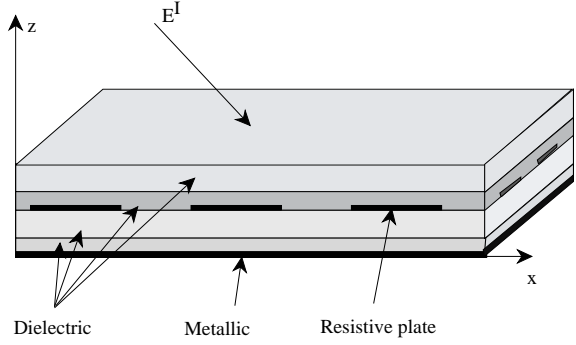}}
\caption{Global geometry of the structure: a grating of
resistive plates between stacks of dielectric media, upon
a perfectly conducting plane.
The incident field propagates in the direction of negative $z$.}
\label{fi:geo}
\end{figure}

We consider a system made of dielectric layers and
biperiodic gratings of resistive conducting plates,
ended by an infinitely conducting plane (or the vacuum),
located at a height $z=0$.
The structure is globally invariant under the discrete
translations of period $(a,b)$ which define the grating,
namely translations of the form $x\to x+ma$, $y\to y+nb$,
$m,n\in\ZZ$, in the $xOy$ plane.
The structure is illuminated by an incident monochromatic field
of the form
$$
\aro{E}^I (x,y,z) = \aro{E}^I \, e^{-i(\omega t-\aro{k}.\aro{r})}\ .
$$
The geometry of the problem is displayed in Fig~\ref{fi:geo}.
We shall generically denote by $E^{(j)}$ and $H^{(j)}$
the electric and magnetic fields in the $j$-th layer $z_j < z < z_{j+1}$,
with electric permittivity $\epsilon_j$; we shall
also use the superscript $+$ or $-$, according to whether the field
propagates in the direction of positive or negative $z$.
From now on, the configuration of Fig.~\ref{fi:geo} will be refered
to as {\em configuration I}.

Alternatively, we shall also consider the same structure, but we remove
the infinitely conducting plane at $z=0$. The latter configuration
will be called {\em configuration II}.
\subsection{Floquet Modes}
Taking into account the global invariance of the problem,
it is natural to introduce the associated Floquet
decompositions. Let $\aro{k} =(k_x,k_y,k_z)$
be the incident wavevector. In a medium of permittivity
$\epsilon$, let us set for all integers $m,n$ 
\beq{fo:alpha.beta}
\left\{
\ba{lll}
&\alpha_0=k_x\sqrt{\epsilon}\ ,
&\alpha_m = \alpha_0 +\ds{\frac{2\pi m}a}\ ,\\
&\beta_0=k_y\sqrt{\epsilon}\ ,
&\beta_n = \beta_0 +\ds{\frac{2\pi n}b}\ ,\\
&t_{mn} = \sqrt{\alpha_m^2 +\beta_n^2}\ ,
&\gamma_{mn}^2 = k^2-t_{mn}^2 \ , 
\ea
\right.
\eeq
where $a$ and $b$ are the grating periods.
For $m,n$ integers, we introduce the corresponding Floquet modes
\beq{fo:floquet.1}
\cE_{mn}^\pm (x,y,z)= \frac1{\sqrt{ab}}e^{i(\alpha_m x+\beta_n y\pm\gamma_{mn}z)}\ ,
\eeq
and the ``planar'' modes
\beq{fo:floquet.2}
\phi_{mn}(x,y) = \frac1{\sqrt{ab}}e^{i(\alpha_m x+\beta_n y)}\ ,
\eeq
which form an orthonormal basis of the space of biperiodic
functions on the plane, with period $(a,b)$.
Then it is well known that such function satisfy Helmholtz's
equation, and that both the electric and the
magnetic fields may be decomposed into those Floquet
modes (see e.g. ~\cite{petit}). Therefore, we write in the
$j$-th layer
\beqa
\aro{E}^{(j)}\!\!\!\!(x,\!y,\!z)\! &=&\!\!
\!\sum_{m,n}\!\! \left(\!\!\aro{e}^{(j)+}_{mn} \!\!\cE_{mn}^+(x,\!y,\!z)
+ \!\! \aro{e}^{(j)-}_{mn}\!\! \cE_{mn}^-(x,y,z)\right)\!,\\
\aro{H}^{(j)}\!\!\!\!(x,\!y,\!z)\! &=& \!\!
\!\sum_{m,n}\!\! \left(\!\!\aro{h}^{(j)+}_{mn} \!\!\cE_{mn}^+(x,\!y,\!z)
+ \!\! \aro{h}^{(j)-}_{mn} \!\!\cE_{mn}^-(x,\!y,\!z)\right)\ ,
\eeqa
where $\aro{e}^{(j)\pm}_{mn}$ and $\aro{h}^{(j)\pm}_{mn}$ denote the
(complex vector) coefficients of the expansion.
The sum over $m,n$ runs theoretically from $-\infty$ to $\infty$. 
In practice it has to be truncated to a finite index $[-M,M]\times [-N,N]$.
We now restrict ourselves to the tangential electric and magnetic fields.
We directly obtain from Maxwell's equations that the following
matrix relations hold
\beq{E2H}
\aro{h}_{mn}^{(j)\pm} =\mp \bK_{mn}^{(j)}\bX\aro{e}_{mn}^{(j)\pm}\ .
\eeq
Here we have introduced the following $2\times 2$ matrices:
\beq{K.X}
\bK_{mn}^{(j)}\! =\! \frac1{\omega\mu\gamma_{mn}}\! \left(\!\ba{cc} k^2-\alpha_m^2&-\alpha_m\beta_n\\
-\alpha_m\beta_n&k^2-\beta_n^2\ea\!\right)\ ,\ 
\bX\! =\!  \left(\!\ba{cc}0&1\\-1&0\ea\!\right)\ .
\eeq
In the following we set $\xi_{mn} = \frac1{\omega\mu\gamma_{mn}}$.

Alternatively, we shall also make use of the expansions with respect to the
planar modes $\phi_{mn}(x,y)$ in~(\ref{fo:floquet.2}), which leads to the
{\em coupled waves}, defined by
\beqa
\aro{E}^{(j)\pm}_{mn}(z) &=&
\aro{e}^{(j)\pm}_{mn}e^{\pm i\gamma_{mn}z}\ ,\\
\aro{H}^{(j)\pm}_{mn}(z) &=&
\aro{h}^{(j)\pm}_{mn}e^{\pm i\gamma_{mn}z}\ .
\eeqa
The propagation of such modes within the corresponding layer is diagonal,
and we have in particular
\beq{prop}
\aro{E}^{(j)\pm}_{mn}(z_{j+1}) = \exp\{\pm i\gamma_{mn}(z_{j+1}-z_j)\}
\aro{E}^{(j)\pm}_{mn}(z_{j})\ .
\eeq
Matching boundary conditions at a dielectric-dielectric
interface is an easy task, since Floquet modes with different
indices are not coupled. Given one such interface between two
dielectric media labeled by $j$,$j+1$, at a height $z=z_j$,
and equating the tangential components of the electric and magnetic
fields, we obtain:
\beq{dioptre}
\left(\!\!\ba{c} \aro{E}_{mn}^{(j)+}\\
\aro{E}_{mn}^{(j)-}\ea\!\!\right)\! =\! C_{mn}^{(j)}\!\!
\left(\!\!\ba{c} \aro{E}_{mn}^{(j+1)+}\\\aro{E}_{mn}^{(j+1)-}\ea\!\!\right)
\!=\!\left(\!\!\ba{cc} \bc&\bc'\\\bc'&\bc\ea\!\!\right)\!\!
\left(\!\!\ba{c} \aro{E}_{mn}^{(j+1)+}\\\aro{E}_{mn}^{(j+1)-}\ea\!\!\right)\ ,
\eeq
where for the sake of simplicity we have suppressed the explicit
dependence on the height $z=z_j$. The matrix elements (which
are themselves $2\times 2$ matrices)
$\bc,\bc'$ are given by
\beqa
\label{bc}
\bc &=\bc^{(j)}_{mn} &= \frac1{2} (\bX)\inv
\left(1+(\bK_{mn}^{(j)})\inv\bK_{mn}^{(j+1)}\right)\bX\ ,\\
\bc' &=\bc'^{(j)}_{mn}&= \frac1{2} (\bX)\inv
\left(1-(\bK_{mn}^{(j)})\inv\bK_{mn}^{(j+1)}\right)\bX\ .
\eeqa

Alternatively, we shall make use of the following $R$-matrices, which read
\beq{Rmat}
\left(\!\!\ba{c} \aro{E}_{mn}^{(j+1)+}\\
\aro{E}_{mn}^{(j)-}\ea\!\!\right)\! =\! R_{mn}^{(j)}\!
\left(\!\!\ba{c} \aro{E}_{mn}^{(j)+}\\\aro{E}_{mn}^{(j+1)-}\ea\!\!\!\right)
\!=\!\left(\!\!\ba{cc} \bt^{++}&\br^{-+}\\\br^{+-}&\bt^{--}\ea\!\!\right)\!\!
\left(\!\!\ba{c} \aro{E}_{mn}^{(j)+}\\\aro{E}_{mn}^{(j+1)-}\ea\!\!\!\right)\ ,
\eeq
and the connection between the two formulations is given by~\cite{Li}
\beq{C2R}
R_{mn}^{j} = \left(\ba{cc} \bc^{-1}&-\bc^{-1}\bc'\\
\bc'\bc^{-1}&\bc-\bc'\bc^{-1}\bc'\ea\right)\ .
\eeq
\subsection{Surface Elements and Boundary Conditions on the Conducting Plates}
Let us now describe the surface currents living on
the conducting plates. We may expand such currents into Floquet modes
\beq{current.0}
\cJ (x,y) = \sum_{m,n}\cJ_{mn}\phi_{mn}(x,y)\ ,
\eeq
and impose the boundary conditions.

\begin{figure}
\epsfxsize=8cm
\centerline{\epsfbox{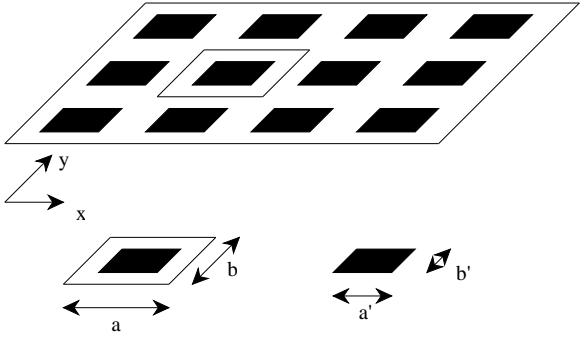}}
\caption{The grating of conducting plates, in the particular
case of rectangular plates.}
\label{fi:rect}
\end{figure}

Several approaches have been proposed for imposing boundary
conditions. Among these, the integral
formulations (e.g. Galerkin methods) are generally considered the most stable.
To implement the Galerkin method, we need to introduce
a family of functions defined on the plates. Let $\aro{\psi}_{pq}(x,y)$
be such a family. If $z_P$ denotes the height of the interface supporting
the conducting plates, we then write, at a height $z=z_P$
\beq{current}
\cJ (x,y) = \sum_{p,q}j_{pq} \aro{\psi}_{pq}(x,y)\ .
\eeq

The boundary conditions rely on three sets of equations.
First, the continuity of the tangential electric fields at all interfaces
\beq{cont.Efield}
\aro{E}^{(j+1)} (x,y,z_j) = \aro{E}^{(j)} (x,y,z_j)\ ,
\eeq
allows one to connect the global electric fields on each side
of the interface. Second, the discontinuity condition for the
tangential magnetic fields:
\beq{imp.bcH}
\aro{H}^{(j+1)} (x,y,z_j)- \aro{H}^{(j)}(x,y,z_j) =  \bX\cJ(x,y,z_j)\ ,
\eeq
explicitely
\beq{imp.bcH.2}
\bK^j\bX\!\!\left(\!\!\aro{E}^{(j)+}_{mn}\!\!\!\!
-\!\aro{E}^{(j)-}_{mn}\!\!\right)\!\! -\!
\bK^{j+1}\bX\!\!\left(\!\!\aro{E}^{(j+1)+}_{mn}\!\!\!\! \!-\!\aro{E}^{(j+1)-}_{mn}\!\!\right)
\!=\!  \bX\cJ_{mn}\ .
\eeq
Finally, the impedance boundary conditions, which read at a height $z=z_P$:
\beq{imp.bcE}
\aro{E}^{(P+1)} (x,y,z_P)= \aro{E}^{(P)}(x,y,z_P) = Z \cJ(x,y,z_P)\ ,
\eeq
(where $\cJ$ vanishes outside the conducting plates)
require a special treatment. It has been observed by several
authors that such conditions cannot be imposed pointwise, because
this leads to unstable systems. Several alternatives have been
proposed and tested (see for example ~\cite{mittra}). The most
stable solutions rely on the use of integral formulations,
obtained by considering either line integrals
of the above equation, or a Galerkin formulation. We limit
ourselves to the latter, which leads to a finite number of
integral equations, obtained by testing Eq.~(\ref{imp.bcE})
against suitably chosen basis functions $\psi_{pq}(x,y)$ (see
Appendix~\ref{se:sur.el} for some possible choices).
\subsection{The Coupled System}
Let us start with the case of {\em configuration I}.
Taking into account the above remarks, we are led to the following formulation.
We denote by $\aro{E}^I$ and $\aro{E}^R$ the incident and reflected
electric fields respectively, and we recall that we have
denoted by $P$ the index of the interface
containing the plates. In order to avoid as much as possible
numerical problems, we limit ourselves to a formulation involving
the so-called $R$-matrix propagation formalism~\cite{Li,Li2}
(see Appendix~\ref{se:R.Mat} for a short account of the method).

Using the $R$-matrix propagation scheme, we can obtain $R$ matrices 
for the stacks below and above $z=z_P$. For example, we obtain a
relation of the form
\beq{coupl.1}
\left(\ba{c}\aro{E}^R_{mn}\\\aro{E}^{(P+1)-}_{mn}\ea\right)=
\left(\ba{cc}\bT^{++}&\bR^{-+}\\\bR^{+-}&\bT^{--}\ea\right)
\left(\ba{c}\aro{E}^{(P+1)+}_{mn}\\\aro{E}^I_{mn}\ea\right)\ ,
\eeq
where the $2\times 2$ matrices $\bT$ and $\bR$ are the stack equivalent
transmission and reflection matrices respectively.
Similarly, the $R$-matrix algorithm below the grating of plates yields
a matrix relation of the form
\beq{coupl.2}
\left(\ba{c}\aro{E}^{(P)+}_{mn}\\\aro{E}^{(0)-}_{mn}\ea\right)=
\left(\ba{cc}\bT'^{++}&\bR'^{-+}\\\bR'^{+-}&\bT'^{--}\ea\right)
\left(\ba{c}-\aro{E}^{(0)-}_{mn}\\\aro{E}^{(P)-}_{mn}\ea\right)\ ,
\eeq
which implies
\beqa
\nonumber
\aro{E}^{(P)+}_{mn}&=&\left(\bR'^{-+} -\bT'^{++}(1-\bT'^{--})^{-1}\bR'^{+-}\right)
\aro{E}^{(P)-}_{mn}\\&=& {\cal N}^{-1}\aro{E}^{(P)-}_{mn}\ .
\label{coupl.3}
\eeqa
The remarkable point with such a formulation is that it only involves
small matrices, since modes with different indices $m,n$ are not coupled.
The only place where coupling between Floquet modes occurs is
at a height $z=z_P$.

\medskip
The case of {\em configuration II} requires only minor modifications.
Eq.~(\ref{coupl.1}) is still valid. For the stack below the
grating of conducting plates, we have to replace Eq.~(\ref{coupl.2}) with
\beq{coupl.2b}
\left(\ba{c}\aro{E}^{(P)+}_{mn}\\\aro{E}^{T}_{mn}\ea\right)=
\left(\ba{cc}\bT'^{++}&\bR'^{-+}\\\bR'^{+-}&\bT'^{--}\ea\right)
\left(\ba{c}0\\\aro{E}^{(P)-}_{mn}\ea\right)\ ,
\eeq
where $\aro{E}^{T}_{mn}$ are the Floquet coefficients of
the transmitted field. Therefore, Eq.~(\ref{coupl.3}) is to be replaced with
\beq{coupl.3b}
\aro{E}^{(P)+}_{mn}= \bR'^{+-} \aro{E}^{(P)-}_{mn}=
{\cal N}^{-1}\aro{E}^{(P)-}_{mn}\ .
\eeq
The rest of the formalism is unchanged.
\section{RESOLUTION AND NUMERICAL RESULTS}
\label{se:num}
\subsection{Resolution of the Coupled System}
We now consider the practical resolution of the
system we have obtained above. We consider approximations
of the fields with $(2N+1)(2M+1)$ Floquet modes
$\aro{E}_{mn},m=-M,\dots M, n=-N,\dots N$, and approximations
of the currents with $PQ$ surface elements $\aro{\psi}_{pq}$.
The boundary conditions lead to three systems of
equations involving the three sets of unknowns:
$\aro{E}^{(P)+}_{mn}$, $\aro{E}^{(P+1)+}_{mn}$ and $\cJ_{mn}$.
Eliminating $\aro{E}^{(P)+}_{mn}$, we first obtain
\beq{resol.1}
\aro{E}^{(P)+}_{mn} \!\!\!\!=
\left(1+\cN\right)\inv\!\left(\!\left(1+\bR^{+-}_{mn}\right)
\aro{E}^{(P+1)+}_{mn}
\!\!\!\!\!\!+\bT^{--}_{mn}\aro{E}^I_{mn}\!\right)\ .
\eeq
Inserting this result into~(\ref{imp.bcH}), we get
\beq{resol.2}
\aro{E}^{(P+1)+}_{mn} =
\bA_{mn}\inv\left( X\cJ_{mn} + \bB_{mn} \aro{E}^I_{mn}\right)\ ,
\eeq
where we have set
\beqa
\nonumber
\bA_{mn}\! &=& \!\! \bK_{mn}^{(p+1)}\!\bX\left(\!\bR_{mn}^{+-}-1\right)\\ &&-
\!\bK_{mn}^{(p)}\!\bX\left(\cN\!-1\right)\left(\cN\!+1\right)\inv \!
\left(\bR_{mn}^{+-}\!+1\!\right)\!\ ,\\
\label{resol.4}
\bB_{mn}\! &=&\!\!\left(\bK_{mn}^{(p)}\!\bX\left(\cN\!-1\right)
\!\left(\cN\!+1\right)\inv
- \!\!\bK_{mn}^{(p+1)}\!\bX\right)\!\bT_{mn}^{--}\ . 
\eeqa
Eventually, we are led to a system of the form
\beq{resol.5a}
\bU_{mn} \aro{E}^I_{mn} = \bV_{mn}\cJ_{mn}\ ,
\eeq
where the $2\times 2$ matrices $\bU_{mn}$ and $\bV_{mn}$ are defined by:
\beqa
\label{resol.5b}
\bU_{mn} &=& \bT_{mn}^{--} +
\left(1+\bR_{mn}^{+-}\right)\bA_{mn}\inv\bB_{mn}\ ,\\
\bV_{mn} &=& Z-\left(1+\bR_{mn}^{+-}\right)\bA_{mn}\inv\bX\ .
\label{resol.5c}
\eeqa
The system~(\ref{resol.5a}) is to be solved numerically, using a
Galerkin procedure.
Let $\aro{\psi}_{pq}(x,y)$ be a basis of functions defined on
the plate, with appropriate boundary conditions. Using the 
expansion~(\ref{current}), we get
\beq{current.2}
\cJ_{mn} = \sum_{p,q} j_{pq} \aro{\psi}_{pq;mn}\ ,
\eeq
where
\beq{cur.coeff}
\aro{\psi}_{pq;mn} = \langle \aro{\psi}_{pq},\phi_{mn}\rangle
=\int \aro{\psi}_{pq}(x,y)\phi_{mn}^*(x,y) dxdy\ ,
\eeq
and where the star ``$*$" denotes complex conjugation.
Taking the scalar products of equations~(\ref{resol.5a}) with the
basis functions $\aro{\psi}_{pq}(x,y)$, we obtain
a system of the form
\beq{current.3}
\cU_{pq} =\sum_{p',q'} \cV_{pq;p'q'} j_{p'q'}\ ,
\eeq
where $\cU$ is a vector of length $PQ$ and $\cV$ is a
$PQ\times PQ$ matrix given by
\beqa
\cU_{pq} &=& \sum_{m,n}\left(\bU_{mn}\aro{E}^I_{mn}\right)\cdot\aro{\psi}_{pq;mn}^*\ ,\\
\cV_{pq;p'q'}&=&\sum_{m,n} \left(\bV_{mn}\aro{\psi}_{p'q';mn}\right)\cdot\aro{\psi}_{pq;mn}^*\ .
\eeqa
Eq.~(\ref{current.3}) is solved numerically (more details are given
in Appendix~\ref{se:num.det}). Once the current $\cJ$
is known, one recovers directly the fields $\aro{E}^{(P+1)+}_{mn}$
using Eq.~(\ref{resol.2}) and then the reflected field $\aro{E}^R$,
from Eq.~(\ref{coupl.1}).
\subsection{Numerical Results}
Our main goal is to exhibit absorption bands, and to analyze the influence
of some specific parameters on the location of the maximal absorption.
More precisely, we focus on the influence of the resistive impedance $Z$
and the ratio {\em size of resistive plates/period}. In addition, we show
that the location of the absorption band essentially does not depend
on the incidence angle. We work with a TM polarization for the incident field
(in fact the results are weakly dependent on the polarization).

We consider a series of configurations, in which we vary individually
these parameters, in the frequency domain $1GHz-10 GHz$. In all the
figures, we plot the reflectivity (i.e. the ratio of reflected flux by
incident flux) as a function of the incident frequency, and in the
case of {\em configuration II} we also plot the transmittivity (i.e.
the ratio of transmitted flux by incident flux).

We start with the case of square resistive plates of sidelength
$a'$, with variable impedance. The period of the grating is
set to $a=10mm$ in both the $x$ and $y$ directions. The grating is
supported by a dielectric stack of height $z=4mm$ and complex
refractive index $\epsilon = 10 +2 i$, itself
supported by an infinitely conducting plane (a simple case of
{\em configuration I}). Since the resistive plates are in that
case square plates, we use Fourier-type decompositions as
described in Appendix~\ref{se:rect} for the decomposition of the
surface current. The numerical results displayed below have been
obtained using $17\times 17$ Floquet modes and the same number of
Galerkin modes.

\begin{figure}
\epsfxsize=9cm
\centerline{\epsfbox{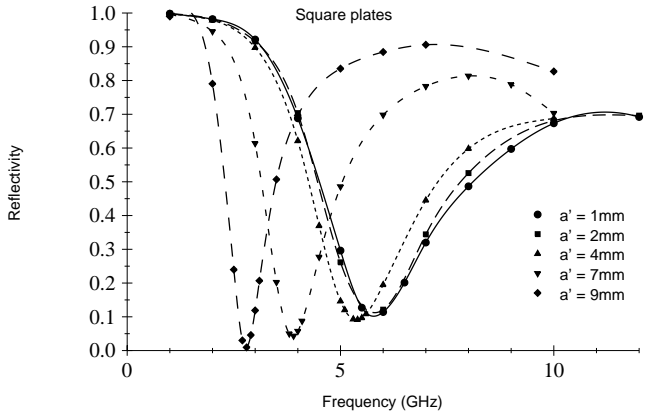}}
\caption{Square plates in {\em configuration I}.
Reflectivity as a function of the incident frequency, for various
values of the plate's size: $a'=1,2,4,7,9mm$. The period $a$ is kept fixed
to $a=10mm$. $\theta=0$ and $Z=0$.}
\label{fi:vary.side}
\end{figure}

We show in Figure~\ref{fi:vary.side} the reflectivity as a function
of the incident frequency, for several values of the ratio $a'/a$.
The computed values are indicated with symbols, and intermediate
values have been obtained using cubic spline interpolation.
In all cases, a significant absorption band is observed. In addition,
the critical frequency (i.e. the frequency at which reflectivity
attains its minimum) decreases as the ratio $a'/a$ increases,
and the width of the absorption band narrows.

In the considered case, the plates are perfectly conducting. We
nevertheless observe a strong absorption in a specific
frequency range. Such a phenomenon is generally coupled with the
excitation of a leaky surface wave. The surface wave may be given
an interpretation in terms of complex poles or zeroes of a scattering
matrix (see~\cite{Newton} for details on the scattering matrix,
and~\cite{neviere} for an analysis
of the role of zeroes and poles).
The poles of the scattering matrix give the propagation constant of
the leaky waves, which propagate along the surface of the
biperiodic grating. The leaky wave is evanescent, as its energy
decreases in the direction normal to the surface of the structure.
The imaginary part of the pole gives the damping of the wave.
The excitation of a leaky wave is a resonance phenomenon at a particular frequency. Figure~\ref{fi:vary.side} shows a spectacular phenomenon.
A highly reflecting capacitive grating (with an important ratio $a'/a$)
can absorb an incident plane wave in totality. Thanks to this absorption
by a leaky surface wave propagating along the grating, we can control the absorption band of a classical Dahlenbach absorber layer which 
consists of
a thick homogeneous lossy layer backed by a metallic plate.
When the ratio $a'/a$ tends to zero we have checked that the minimum of
reflectivity is obtained for the same frequency as in
Figure~\ref{fi:vary.side}.
To adjust the absorption band of the absorber, we can deposit a biperiodic capacitive reflecting grating on the Dahlenbach layer. By doing so, we combine
the properties of the biperiodic grating with those of the lossy layer.
Notice in particular that it is possible to decrease the thickness of the
layer by adding such a biperiodic structure, to obtain the critical
absorption frequency of the initial Dahlenbach structure.

\begin{figure}
\epsfxsize=9cm
\centerline{\epsfbox{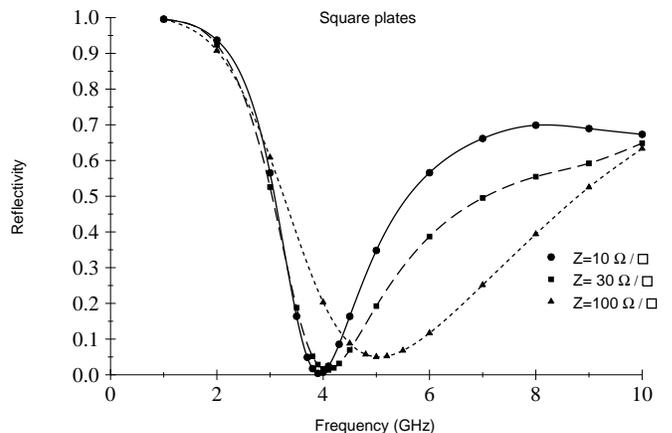}}
\caption{Square plates in {\em configuration I}.
Reflectivity as a function of the incident frequency, for various
values of the plate's impedance: $Z=10,30,100 \Omega/\Box$.
The period $a$ and the plate's size $a'$ are kept fixed
$a=10mm$ and $a'=7mm$. $\theta=0$.}
\label{fi:vary.imp}
\end{figure}

We show in Figure~\ref{fi:vary.imp} 
the reflectivity as a function of the frequency of the incident
beam, for several values of the impedance $Z$.
The configuration corresponds to the case of Fig.~\ref{fi:vary.side}
with $a'=7mm$,
and a significant minimum in the reflectivity is observed for a
certain value of the frequency. This critical value is seen to be
an increasing function of the impedance of the conducting plates.

In Figure~\ref{fi:vary.imp}, the patches of the grating are not perfectly conducting any more. In that case, the absorption frequency and the
bandwidth increase with the resistivity of the patches.
To obtain a required absorption band, it is therefore possible to
combine the effects of the geometry (here the ratio $a'/a$) and the
effect of the conductivity. This provides extra flexibility to
the filter design.

\begin{figure}
\epsfxsize=9cm
\centerline{\epsfbox{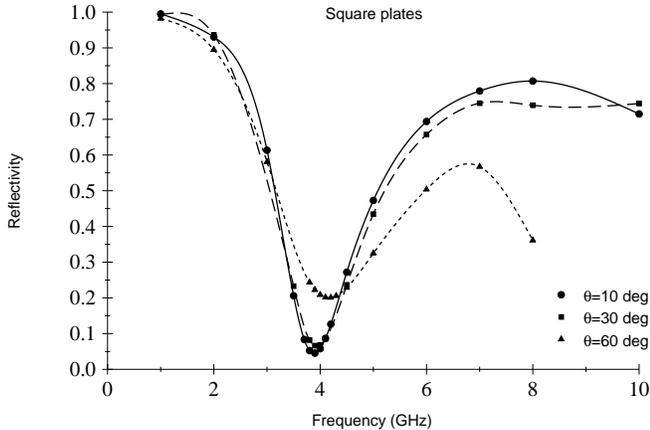}}
\caption{Square plates in {\em configuration I}.
Reflectivity as a function of the incident frequency, for various
values of incidence angle: $\theta=10,30,60\ deg$.
The period $a$ and the plate's size $a'$ are kept fixed
$a=10mm$ and $a'=7mm$. $Z=0$.}
\label{fi:vary.ang}
\end{figure}

We show in Figure~\ref{fi:vary.ang}
the reflectivity as a function of the frequency of the incident
beam, for several values of the incidence
angle $\theta$, for the same configuration as before, i.e. a configuration exhibiting a well defined absorption band.
These results (and other tests of intermediate incidence angles,
are not reproduced here to simplify the plot) show that the critical
frequency value depends very weakly on the incidence angle
(at least for angles smaller than 45 deg).

\bigskip
The same computations have been performed with resistive plates
of various shapes. We display here the results obtained when the
square resistive plates in Fig.~\ref{fi:vary.side}
are replaced with cross-shaped ones, of the same size.
By this we mean that the crosses lie within a square of the sidelength
$a'$, and are made of five identical squares of sidelength $a'/3$.
For this case, we used the surface elements described in
Appendix~\ref{se:FE}, and as before we take
$17\times 17$ Floquet modes, and the same number of Galerkin modes.

\begin{figure}
\epsfxsize=9cm
\centerline{\epsfbox{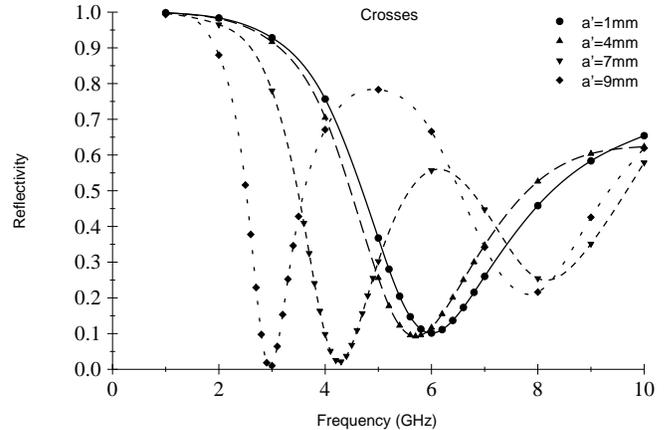}}
\caption{Cross-shaped plates in {\em configuration I}.
Reflectivity as a function of the incident frequency, for various
values of the plate's size: $a'=1,2,4,7,9mm$. The period $a$ is kept fixed
to $a=10mm$. $\theta=0$ and $Z=0$.}
\label{fi:vary.side.2}
\end{figure}

\begin{figure}
\epsfxsize=9cm
\centerline{\epsfbox{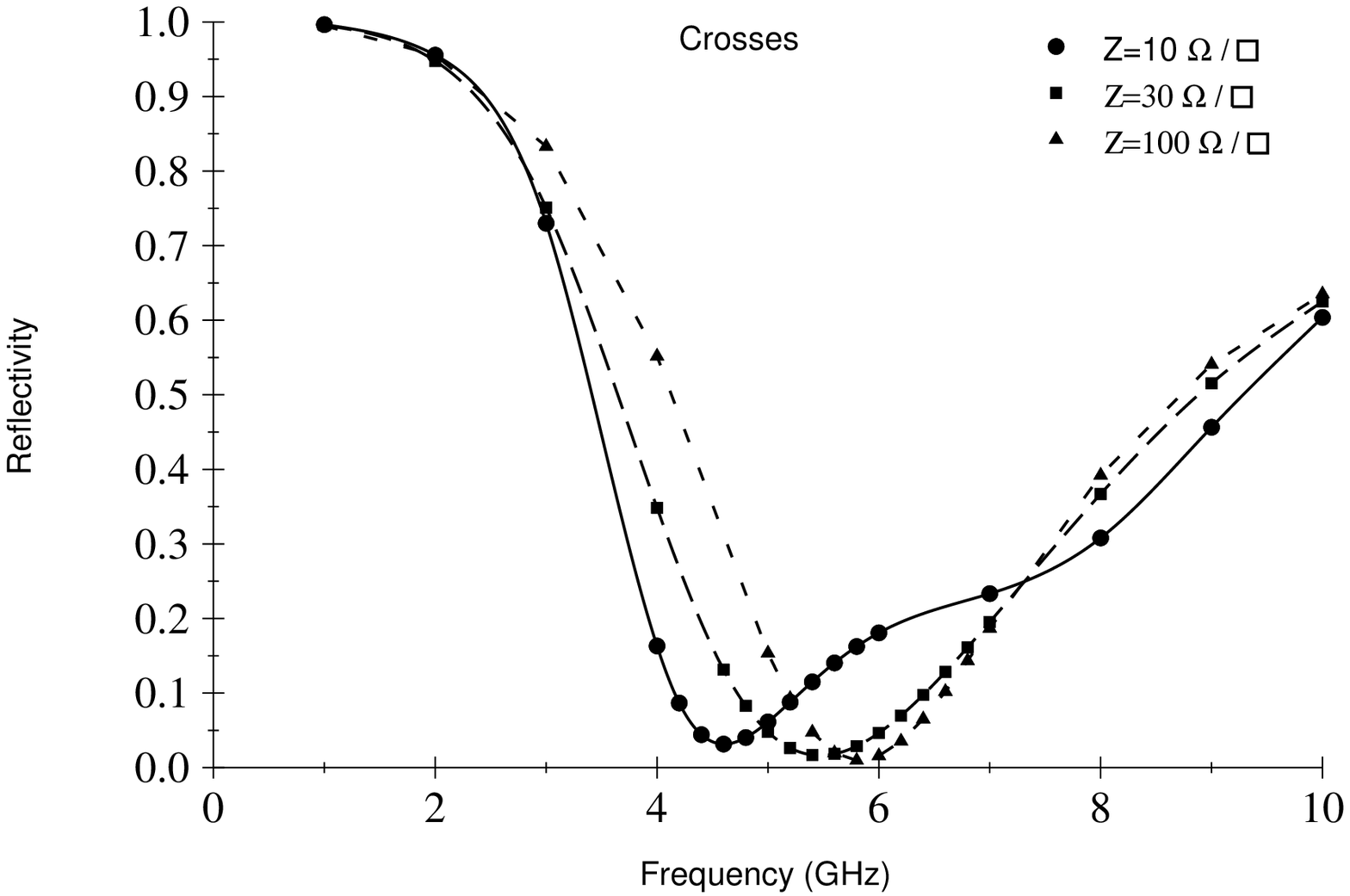}}
\caption{Cross-shaped plates in {\em configuration I}.
Reflectivity as a function of the incident frequency, for various
values of the plate's impedance: $Z=10,30,100 \Omega/\Box$.
The period $a$ and the plate's size $a'$ are kept fixed
$a=10mm$ and $a'=7mm$. $\theta=0$.}
\label{fi:vary.imp.2}
\end{figure}

The numerical results, displayed in Figures~\ref{fi:vary.side.2}
and~\ref{fi:vary.imp.2} show a similar behavior to the previous case: a
well defined absorption band is clearly seen, and the critical
frequency again depends on the ratio $a'/a$ and on the impedance $Z$.
Again, the location of the absorption band depends
only weakly on the incidence angle (the numerical results, not
given here, are very similar to those displayed in
Fig.~\ref{fi:vary.ang}). The only significant difference
which may be observed is a broadening of the absorption band
in the case of cross-shaped plates, and a second minimum occurs
for large $a'$.

\begin{figure}
\epsfxsize=9cm
\centerline{\epsfbox{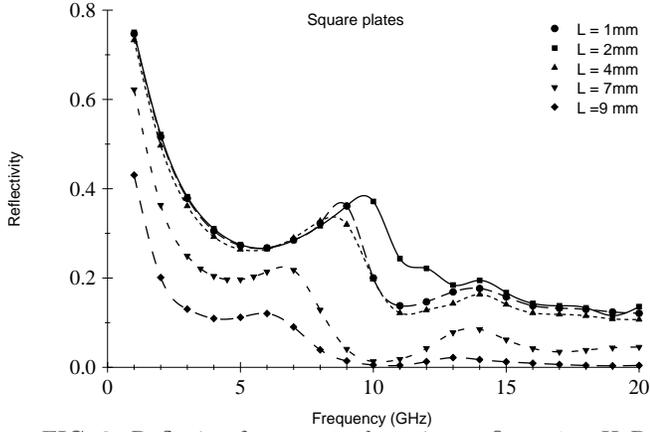}}
\caption{Reflexion for square plates in {\em configuration II}.
Reflectivity as a function of the incident frequency, for various
values of the plate's size: $a'=1,2,4,7,9mm$. The period $a$ is kept fixed
to $a=10mm$. $\theta=0$ and $Z=0$.}
\label{fi:config.2}
\end{figure}

\begin{figure}
\epsfxsize=9cm
\centerline{\epsfbox{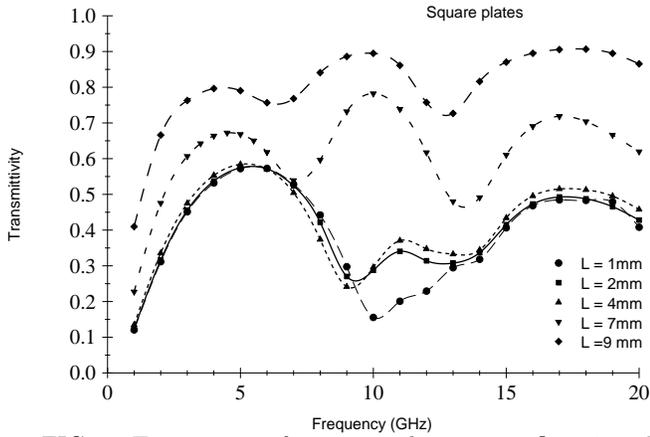}}
\caption{Transmission for square plates in {\em configuration II}.
Reflectivity as a function of the incident frequency, for various
values of the plate's size: $a'=1,2,4,7,9mm$. The period $a$ is kept fixed
to $a=10mm$. $\theta=0$ and $Z=0$.}
\label{fi:config.2b}
\end{figure}

\bigskip
Similar computations have been made with {\em configuration II}. We
display in Fig.~\ref{fi:config.2} (reflexion) and Fig.~\ref{fi:config.2b}
(transmission) the results obtained with systems identical to those
considered in Figures~\ref{fi:vary.side}-\ref{fi:vary.ang}. We
observe that in such a configuration, the reflexion is small and
almost constant above $5 Ghz$, it increases slightly with $a'$.
The transmittivity shows maximums at frequencies corresponding to
the minimums in {\em configuration I}.

These two figures show the importance of the conducting plane at $z=0$.
The well-defined absorption band appears only in that case. The excitation
of the leaky wave and the corresponding absorption occurs only for
structures ended by a conducting plane.

\bigskip
Next, we consider a second system (in {\em configuration I}), in which
the resistive plates are located upon a double layer of dielectrics.
The first dielectric (upon which the plates are located) has electric
permittivity $\epsilon = 5$, and the second layer has electric
permittivity $\epsilon = 15 + i 18\sigma /\nu$ with a
frequency dependent imaginary part. Here the constant
$\sigma$ is set to $\sigma = 10 s/m$, and the frequency $\nu$
is expressed in $GHz$.

\begin{figure}
\epsfxsize=9cm
\centerline{\epsfbox{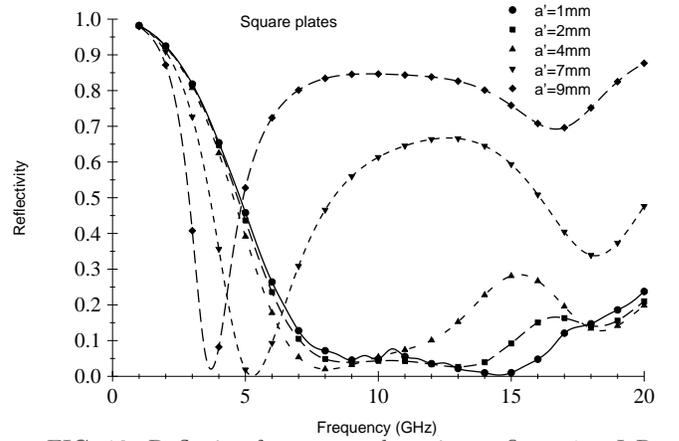}}
\caption{Reflexion for square plates in {\em configuration I}.
Reflectivity as a function of the incident frequency, for various
values of the plate's size: $a'=1,2,4,7,9mm$. The period $a$ is kept fixed
to $a=10mm$. $\theta=0$ and $Z=0$.}
\label{fi:fig4}
\end{figure}

\begin{figure}
\epsfxsize=9cm
\centerline{\epsfbox{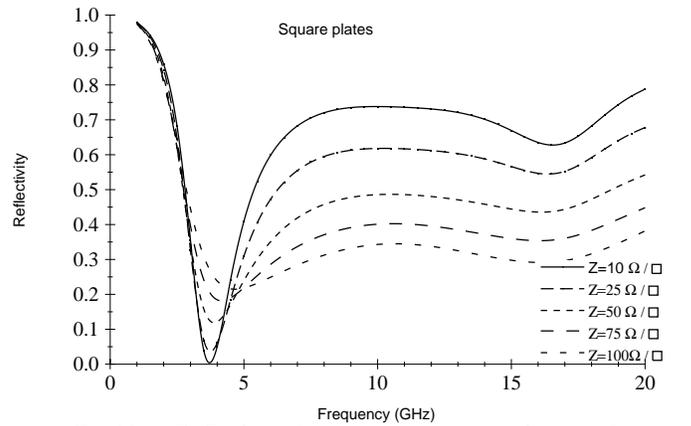}}
\caption{Reflexion for square plates in {\em configuration I}.
Reflectivity as a function of the incident frequency, for various
values of the resistive impedance $Z$: $Z=10,25,50,75,100 \Omega/\Box$.
The period $a$ is kept fixed
to $a=10mm$. $a'=7mm$ and $\theta=0$.}
\label{fi:imped}
\end{figure}

The results are displayed in Figures~\ref{fi:fig4} and~\ref{fi:imped}.
As before, an absorption band is clearly seen on Figure~\ref{fi:fig4},
when $a'$ is above 4mm, whose critical frequency and bandwidth decrease
as the sidelength of the plates increases.
In addition, for small plates, the reflectivity has a constant
behavior close to zero above $8 GHz$.
Figure~\ref{fi:imped} shows that in such a configuration, the critical 
frequency depends weakly on the value of the impedance, but the
bandwidth is an increasing function of the impedance.

\section{CONCLUSIONS AND PERSPECTIVES}
\label{se:concl}
We have studied and described a series of configurations involving dielectric
stacks and arrays or resistive plates which produce well-defined
absorption bands, with controllable absorption frequency.
The critical frequency has been shown to be strongly influenced by the
ratio period/plate-size, which therefore provides a good
control parameter. The impedance of the resistive plates
has been shown to allow the control of the critical frequency.

Our approach is based on a Floquet (or Rayleigh) development of
the electromagnetic fields within the different layers of the structure,
and a Galerkin approximation of the surface currents. Multilayers more
complex than the ones we considered here may be described by the
formalism of this paper as well.

In light of the numerical experiments we have performed, it is possible
to combine the different parameters (namely the ratio $a'/a$, the geometry
of the patches and the conductivity of the patch material) to obtain
optimized absorbing structures from a quite standard biperiodic grating.
The use of absorption by a leaky surface wave can improve a classical 
Dahlenbach structure.

\acknowledgements
We thank P. Chiappetta and A. Grossmann for stimulating discussions.
C. Bourrely would like to thank Prof. E. Leader for his invitation at
Birkbeck College.
C. Ordenovic is supported by Thomson CSF-Optronique
and the French government under contract CIFRE number 400/95.

\appendix
\section{THE SURFACE ELEMENTS}
\label{se:sur.el}
Depending on the geometry of the conducting plates, several
different bases of surface elements may be used. In all cases,
the finite number of basis functions we are forced to consider
limits the precision of the approximation of the current.
\subsection{Rectangular Plates}
\label{se:rect}
To start with, we consider the case of rectangular plates, as shown
in Fig.~\ref{fi:rect} above.
In such cases, the best choice for surface elements is provided
by a Fourier basis: we set
\beqa
\vec{\psi}^{TE}_{pq}(x,y) & = &~~{p\pi \over a'}\sin{{p\pi \over a'} [ x + 
\hf a'  ]} \cos{{q\pi \over b'} [ y +\hf b' ]} \vec{e}_x 
\nonumber \\ 
& & + {q\pi \over b'}\cos{{p\pi \over a'} [ x + 
\hf a' ]} \sin{{q\pi \over b'} [ y +\hf b' ]} \vec{e}_y\ ,
\label{d5}
\eeqa
\beqa
\vec{\psi}^{TM}_{pq}(x,y) & = &~~{q\pi \over b'}\sin{{p\pi \over a'} [ x + 
\hf a'  ]} \cos{{q\pi \over b'} [ y +\hf b' ]} \vec{e}_x 
\nonumber \\ 
& & - {p\pi \over a'}\cos{{p\pi \over a'} [ x + 
\hf a' ]} \sin{{q\pi \over b'} [ y +\hf b' ]} \vec{e}_y\ .
\label{d6}
\eeqa
Therefore, the Floquet modes of the surface current may be written as
\beq{d3}
\cJ_{mn} =  \sum_{p=0}^{P-1} \sum_{q=0}^{Q-1}
\left( j_{pq}^{TM} \aro{\psi}^{TM}_{pq,mn} +
 j_{pq}^{TE}\aro{\psi}^{TE}_{pq,mn}\right)\ ,
\eeq
and the scalar products
\beqa
\aro{\psi}^{TE}_{pq,mn} &=& \la \vec{\psi}^{TE}_{pq},\phi_{mn}\ra\ ,\\
\aro{\psi}^{TM}_{pq,mn} &=& \la \vec{\psi}^{TM}_{pq},\phi_{mn}\ra\ ,
\eeqa
may be computed analytically.

For other special geometries, such as disks or elongated disks, it
is possible to design appropriate basis functions to describe the
current density on the resistive plates (in the case of disks,
such basis functions are linear combinations of Bessel functions).
However, it is also desirable to have basis functions which can
describe arbitrary geometries. This is the purpose of the surface
elements described in the next subsection.
\subsection{Arbitrary Plates}
\label{se:FE}
For conducting plates with arbitrary geometry, we are forced to use
``all purpose'' basis functions, which we shall call surface elements.
Such basis functions have been considered by several authors under
the name of {\em rooftop functions}. It follows from the analysis 
in~\cite{mittra} that rooftop functions often provide faster and better conditioned numerical schemes than classical alternatives (the
so-called surface-patch and triangular patch functions).
The first step for the construction of
such surface elements is a discretization of the plate. For the sake of
simplicity, we restrict to a uniform square discretization, with period
$\tau$. Consider the characteristic function
\beq{chi}
\chi(x) = \left\{\ba{ll}
1&\hbox{ if } 0\le x\le \tau\\
0&\hbox{ elsewhere} \ea\right.
\eeq
and the Schauder function
\beq{lambda}
\Lambda (x) = \left\{\ba{ll}
1+\frac{x}\tau &\hbox{ if } -\tau\le x\le 0\\
1-\frac{x}\tau &\hbox{ if } 0\le x\le \tau\\
0&\hbox{ elsewhere} \ea\right.
\eeq
Then set
\beq{psi.x}
\psi^x_{pq} (x,y) = \chi(x-p\tau)\Lambda(y-q\tau)\ ,
\eeq
\beq{psi.y}
\psi^y_{pq} (x,y) = \Lambda(x-p\tau)\chi(y-q\tau)\ ,
\eeq
and finally
\beq{psi.vec}
\aro{\psi}_{pq}(x,y) = \psi_{pq}^x\vec{e}_x +\psi_{pq}^y\vec{e}_y\ .
\eeq
The surface elements we consider will be those functions
$\psi^x_{pq} (x,y)$ and $\psi^y_{pq} (x,y)$ such that their
support is completely included in the support of the plate.
Clearly, the smaller $\tau$ the better is the approximation of
the current, but the higher the complexity of the numerical problem.
\section{$R$-MATRIX PROPAGATION}
\label{se:R.Mat}
We describe briefly the $R$-matrix propagation scheme as we used it
in our simulations. Clearly, the simplest approach amounts to consider
the direct product of the $C$ matrices given in Eq.~(\ref{dioptre}),
which yields directly a $C$ matrix for the whole structure.
As stressed by various authors, such a scheme turns out to
become rapidly unstable as the depth of the structure grows.

\begin{figure}
\epsfxsize=6cm
\centerline{\epsfbox{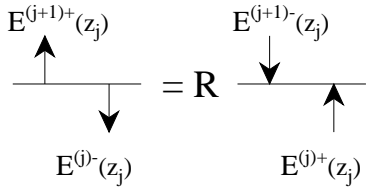}}
\caption{Illustration of the $R$-matrix propagation algorithm:
the role of an interface $R$-matrix.}
\label{fi:stack}
\end{figure}

Let us consider a multilayered medium with interfaces
at heights $z_p,\dots,z_F$, and assume that we are given an
interface $R$-matrix of the form given in Eq.~(\ref{Rmat}).
Then, one easily verifies that 
\beq{app.1}
\left(\ba{c}\aro{E}^{(j+1)+}(z_j)\\\aro{E}^{(j)-}(z_{j-1})\ea\right) =
\left(\ba{cc}
\tbt^{++}&\tbr^{-+}\\\tbr^{-+}&\tbt^{++}
\ea\right)
\left(\ba{c}\aro{E}^{(j)+}(z_{j-1})\\\aro{E}^{(j+1)-}(z_{j})\ea\right)\ ,
\eeq
where we have set
\beq{app.2}
\left\{\ba{l} \tbt^{++}=\bt^{++}L_j\, ;\ \tbt^{--}=\bt^{--}L_j\, ;\\ 
\tbr^{++}=\br^{+-}L_j^2\, ;\ \tbr^{-+}=\br^{-+}\ ,\\
L_j=\exp\{i\gamma_{mn}(z_{j+1}-z_j)\}\ .
\ea\right.
\eeq

\begin{figure}
\epsfxsize=6cm
\centerline{\epsfbox{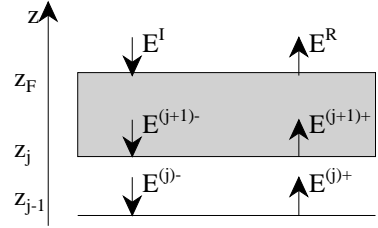}}
\caption{Illustration of the $R$-matrix propagation algorithm:
the stack $R$-matrix.}
\label{fi:stack2}
\end{figure}

Suppose now that we are given a stack $R$-matrix for the stack
$[j+1,F]$:
\beq{app.3}
\left(\ba{c}\aro{E}^{R}\\\aro{E}^{(j)-}\ea\right) =
\left(\ba{cc}
\bT^{++}&\bR^{-+}\\\bR^{-+}&\bT^{++}
\ea\right)
\left(\ba{c}\aro{E}^{(j)+}\\\aro{E}^{I}\ea\right)\ ,
\eeq
where we set by default $\aro{E}^{(j)\pm} =
\aro{E}^{(j)\pm} (z_{j-1})$ for the sake of simplicity.
From Eqs.~(\ref{app.1}) and~(\ref{app.3}), little algebra
gives the expression of the coefficients of the stack
matrix for the stack $[j+1,F]$:
\beqa
\label{app.4}
\nonumber
\aro{E}^R(z_F) &=&\bT^{++}(1\!\!-\bR^{+-}\tbr^{-+})\inv\tbt^{++}
\aro{E}^{(j)+}\!\!\!\!\!(z_{j-1})\\
\nonumber
&+&\left(\bR^{-+}\!\! +\bT^{++}(1\!\!-\bR^{+-}\tbr^{-+})\inv\tbr^{-+}\bT^{--}
\right)\aro{E}^I(z_F)\ ,\\
\nonumber
\aro{E}^{(j)-}\!\!\!\!\!(z_{j-1})&=& \left(\tbr^{+-}\!\! +\!
\tbt^{--}(1\!\!-\bR^{+-}\tbr^{-+})\inv\tbr^{-+}\tbt^{++}\right)
\aro{E}^{(j)+}\!\!\!\!\!(z_{j-1})\\
\nonumber
&+&\tbt^{--}(1\!\!-\bR^{+-}\tbr^{-+})\inv\bT^{--}\aro{E}^I(z_F)\ .
\eeqa
The above equations provide a simple iterative algorithm for computing
the global $R$-matrix for the stacks $[z_P,z_F]$ and $[z_0,z_P]$.
This algorithm is known as the {\em $R$-matrix propagation algorithm},
and has been analyzed by various authors. We refer to~\cite{Li,Li2,Ne,PA}
for more details.

\section{NUMERICAL ASPECTS}
\label{se:num.det}
We give here more details on the numerical methods used to solve the
complete problem. As stressed before, most of the matrices used in
the scheme are $2\times 2$ matrices, which are easy to handle. In
addition, the use of $R$-matrix propagation algorithm prevents
us from developing numerical instabilities when computing
products of such matrices.

The main part of CPU is used for solving Eq.~(\ref{current.3}).
Several methods have been tested for that problem (which has
also been studied by various authors). The numerical results
presented here have been obtained by using an inversion
method based on $LU$-decomposition, with left and rigth
equilibrations of the matrix. A fortran implementation of
such a method is available in the LAPACK library (see~\cite{lapack}).
Alternative methods may be found in the literature, such as
(complex) biconjugate gradient methods or FFT-based methods.
\end{document}